\documentclass{svjour3}
\smartqed  % flush right qed marks, e.g. at end of proof
\usepackage{graphicx}

\journalname{Few-Body Systems}
\begin{document}

\title{Renormalization group invariance in pionless effective field theory for the $NN$ system
\footnote{Presented at the 21st European Conference on Few-Body Problems in Physics,
Salamanca, Spain, 30 August - 3 September 2010} }

\author{V. S. Tim\'oteo         \and
        S. Szpigel \and  F. O. Dur\~aes
}

\institute{V. S. Tim\'oteo \at
              Faculdade de Tecnologia, Universidade Estadual de Campinas - UNICAMP \\
13484-332, Limeira, SP, Brasil
           \and
           S. Szpigel and  F. O. Dur\~aes \at
              Centro de Ci\^encias e Humanidades, Universidade Presbiteriana Mackenzie \\
01302-907, S\~ao Paulo, SP, Brasil
}

\date{Received: date / Accepted: date}
% The correct dates will be entered by the editor

\maketitle

\begin{abstract}
We consider the $NN$ interaction in pionless effective field theory (EFT) up to next-to-next-to-leading order (NNLO) and use a recursive subtractive renormalization scheme to describe $NN$ scattering in the $^1S_0$ channel. We fix the strengths of the contact interactions at a reference scale, chosen to be the one that provides the best fit for the phase-shifts, and then slide the renormalization scale by evolving the driving terms of the subtracted Lippmann-Schwinger equation through a non-relativistic Callan-Symanzik equation. The results show that such a systematic renormalization scheme with multiple subtractions is fully renormalization group invariant.

\end{abstract}

\section{Introduction}
\label{intro}

The standard method for the non-perturbative renormalization of the $NN$ interaction in the context of Weinberg's approach to chiral effective field theory (ChEFT) consists of two steps \cite{epelbaum}. First, one solves the regularized Lippmann-Schwinger (LS) equation for the scattering amplitude with the $NN$ potential truncated at a given order in the chiral expansion. The most common scheme used to regularize the LS equation is to introduce a sharp or smooth regularizing function that suppresses the contributions from the potential matrix elements for momenta larger than a given cutoff scale, thus eliminating the ultraviolet divergences in the momentum integrals. Then, one needs to determine the strengths of the contact interactions, the so called low-energy constants (LECs), by fitting a set of low-energy scattering data. Once the LECs are fixed for a given cutoff, the LS equation can be solved to evaluate other observables.

Such a procedure, motivated by Wilson's renormalization group \cite{wilson1,wilson2}, relies on the fundamental premise of EFT that physics at low-energy/long-distance scales is insensitive with respect to the details at high-energy/short-distance scales \cite{lepage}, i.e. the relevant high-energy/short-distance effects for describing the low-energy observables can be captured in the cutoff-dependent LECs. The NN interaction can be considered properly renormalized when the calculated observables are independent of the cutoff scale within the range of validity of the ChEFT or involves a small residual cutoff dependence due to the truncation of the chiral expansion. In the language of Wilson's renormalization group, this means that the LECs must run with the cutoff scale in such a way that the scattering amplitude becomes renormalization group invariant (RGI).

An alternative approach is the subtracted kernel method (SKM) \cite{npa99,plb00,hepph01,plb05,lc09,fb19,iwara09b,aop2010} in which, instead of using a cutoff function, the LS equation is regularized by performing subtractions in the kernel. An advantage of the SKM approach is that it can be recursively extended to any derivative order of the contact interactions. In this work we apply the SKM approach to renormalize $NN$ interactions in pionless EFT, which consist of Dirac-delta plus derivative contact interactions. In this case, the SKM procedure can be performed analytically. In ChEFT, where both contact and pion-exchange interactions are included, the procedure must be performed numerically.

\section{SKM in Pionless EFT}
\label{piless}

Consider the scattering of two nucleons in the $^1S_0$ channel interacting through the pionless EFT potential. In a partial-wave relative momentum space basis, the matrix elements of such a potential up to NNLO are given by
\begin{eqnarray}
V_{\mu}(p,p') = C_0(\mu) + C_2(\mu)~(p^2+p'^2) + C_4(\mu)~p^2 ~ p'^2 + C'_4(\mu)~(p^4+p'^4)\; .
\label{contact}
\end{eqnarray}
\noindent
In the SKM approach a regularized and renormalized LS equation for the $T$-matrix is derived through an iterative process involving multiple subtractions in kernel at a certain energy scale $-\mu^2$.
%For a general number of subtractions $n$, we define a $n$-fold subtracted kernel LS equation, given by
%%
%\begin{equation}
%T(E) = V^{(n)}(E;-\mu^2) + V^{(n)}(E;-\mu^2)~G_{n}^{+}(E;-\mu^2)~T(E) \; ,
%\label{LSn}
%\end{equation}
%%
%\noindent
%where $k=\sqrt{E}$ is the on-shell momentum in the center-of-mass frame. The $n$-fold subtracted Green's function $G_{n}^{+}(E;-\mu^2)$ is defined by
%%
%\begin{equation}
%G_{n}^{+}(E;-\mu^2) \equiv \left[(-\mu^2-E)~ G_{0}^{+}(-\mu^2) \right]^{n}~G_{0}^{+}(E)  \; .
%\label{Gn}
%\end{equation}
%%
%where $G_{0}^{+}(E)$ is the free Green's function.
Here, for convenience, we implement the SKM procedure using the $K$-matrix instead of the $T$-matrix. The driving term $V_\mu^{(n)}(E)$ is recursively computed through an iterative procedure, starting from the ansatz $V^{(1)}_{\mu}(p,p') = C_0(\mu)$.

%For the LO interaction, we solve the LS equation for the $K$-matrix with one subtraction,
%%
%\begin{eqnarray}
%K_{1}(p,p') = V^{(1)}_{\mu}(p,p') + \frac 2 \pi \mathcal{P} \int_0^\infty dq~q^2
%\left(\frac{\mu^2+k^2}{\mu^2+q^2}\right) V^{(1)}_{\mu}(p,q) \frac{1}{k^2-q^2}K_{1}(q,p') \; .
%\label{subk1}
%\end{eqnarray}
%%
%
For the NLO interaction, we calculate $V^{(2)}$ from $V^{(1)}$ through the equation
\begin{eqnarray}
V^{(2)}_{\mu}(p,p') = V^{(1)}_{\mu}(p,p') + \frac 2 \pi \int_0^\infty dq~q^2 ~ V^{(1)}_{\mu}(p,q)
\frac{(\mu^2+k^2)^{1}}{(\mu^2+q^2)^2}V^{(2)}_{\mu}(q,p') \; .
\label{v2}
\end{eqnarray}
\noindent
Then, we calculate $V^{(3)}$ from $V^{(2)}$,
\begin{eqnarray}
V^{(3)}_{\mu}(p,p') = V^{(2)}_{\mu}(p,p') + \frac 2 \pi \int_0^\infty dq~q^2 ~V^{(2)}_{\mu}(p,q)
\frac{(\mu^2+k^2)^{2}}{(\mu^2+q^2)^3}V^{(3)}_{\mu}(q,p) \; ,
\label{v3}
\end{eqnarray}
and add the NLO contact interaction $V^{(3)}_{\mu,{\rm cont}}(p,p')= ~C_2(\mu)~(p^2+p'^2)$.

%The resulting LS equation with three subtractions is given by
%%
%\begin{eqnarray}
%K_{3}(p,p') = V^{(3)}_{\mu}(p,p') + \frac 2 \pi \mathcal{P} \int_0^\infty dq~q^2
%\left(\frac{\mu^2+k^2}{\mu^2+q^2}\right)^3 V^{(3)}_{\mu}(p,q) \frac{1}{k^2-q^2}K_{3}(q,p') \; .
%\label{subk3}
%\end{eqnarray}
%%
%
For the NNLO interaction, we calculate $V^{(4)}$ from $V^{(3)}$,
\begin{eqnarray}
V^{(4)}_{\mu}(p,p') = V^{(3)}_{\mu}(p,p';k^2) + \frac 2 \pi \int_0^\infty dq~q^2 ~ V^{(3)}_{\mu}(p,q)
\frac{(\mu^2+k^2)^{3}}{(\mu^2+q^2)^4}V^{(4)}_{\mu}(q,p') \; ,
\label{v4}
\end{eqnarray}
and add the NNLO contact interaction $V^{(4)}_{\mu,{\rm cont}}(p,p')= ~C_4(\mu)~p^2 ~ p'^2$ (we neglect the last term in Eq.~(\ref{contact}), since it does not significantly change the results).

The LS equation with $n$ subtractions is given by
\begin{eqnarray}
K_{n}(p,p') = V^{(n)}_{\mu}(p,p') + \frac 2 \pi \mathcal{P} \int_0^\infty dq q^2
\left(\frac{\mu^2+k^2}{\mu^2+q^2}\right)^n V^{(n)}_{\mu}(p,q) \frac{1}{k^2-q^2}K_{n}(q,p') \; .
\label{subk4}
\end{eqnarray}

One important feature of the SKM approach is that, by requiring the $K$-matrix to be invariant with respect to the scale $\mu$, a renormalization group flow equation can be obtained for the recursive driving terms in the form of a non-relativistic Callan-Symanzik equation (NRCS). Here we consider the renormalization group invariance of the NLO interaction, which requires three subtractions. The NRCS equation for the driving term $V^{(3)}_{\mu}(p,q)$ reads
\begin{eqnarray}
\frac{\partial}{\partial\mu^2}V^{(3)}_{\mu}(p,p') = \frac 2 \pi \int_0^\infty dq~q^2~
V^{(3)}_{\mu}(p,q) \frac{3(\mu^2+k^2)^{2}}{(\mu^2+q^2)^{4}}V^{(3)}_{\mu}(q,p') \; .
\label{cse3}
\end{eqnarray}
\noindent
with the boundary condition given by $V^{(3)}_{\mu}|_{\mu \rightarrow {\bar \mu}}= V^{(3)}_{{\bar \mu}}$ imposed at some reference scale $\bar \mu$. Thus, once the renormalized strengths are fixed at the reference scale ${\bar \mu}$ to fit the observables used as physical input, the scale $\mu$ can be changed without modifying the results for the calculated observables.

\section{Numerical Results}
\label{results}

\begin{figure}[t]
\begin{center}
\includegraphics[scale=0.5]{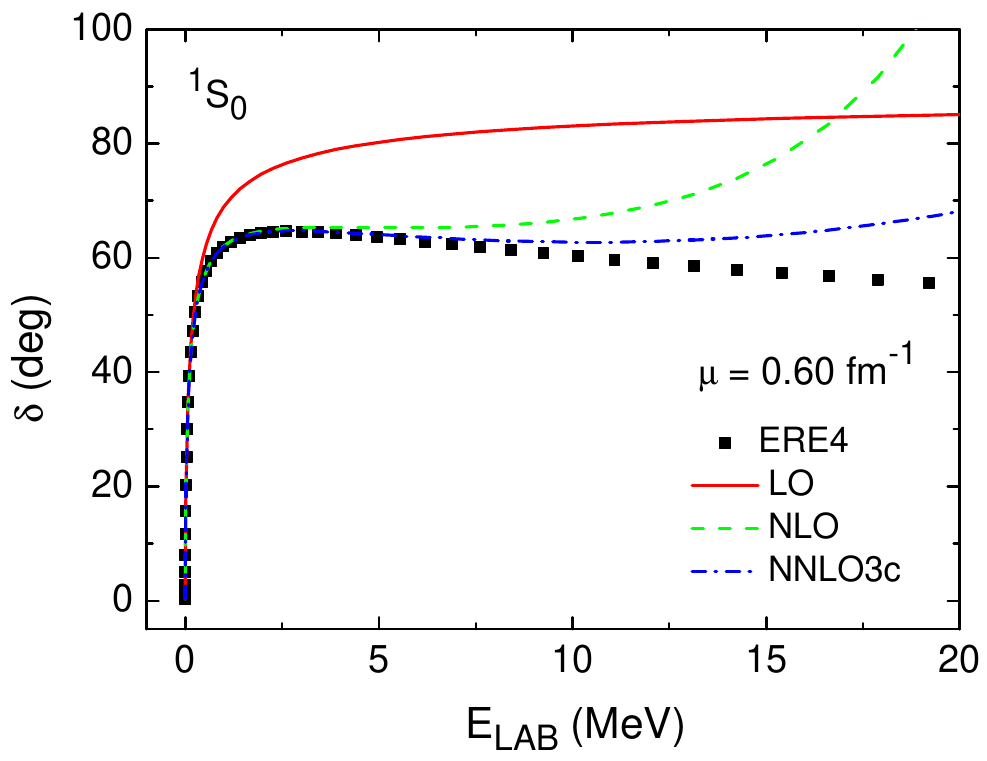}
\hspace{0.7cm}
\includegraphics[scale=0.5]{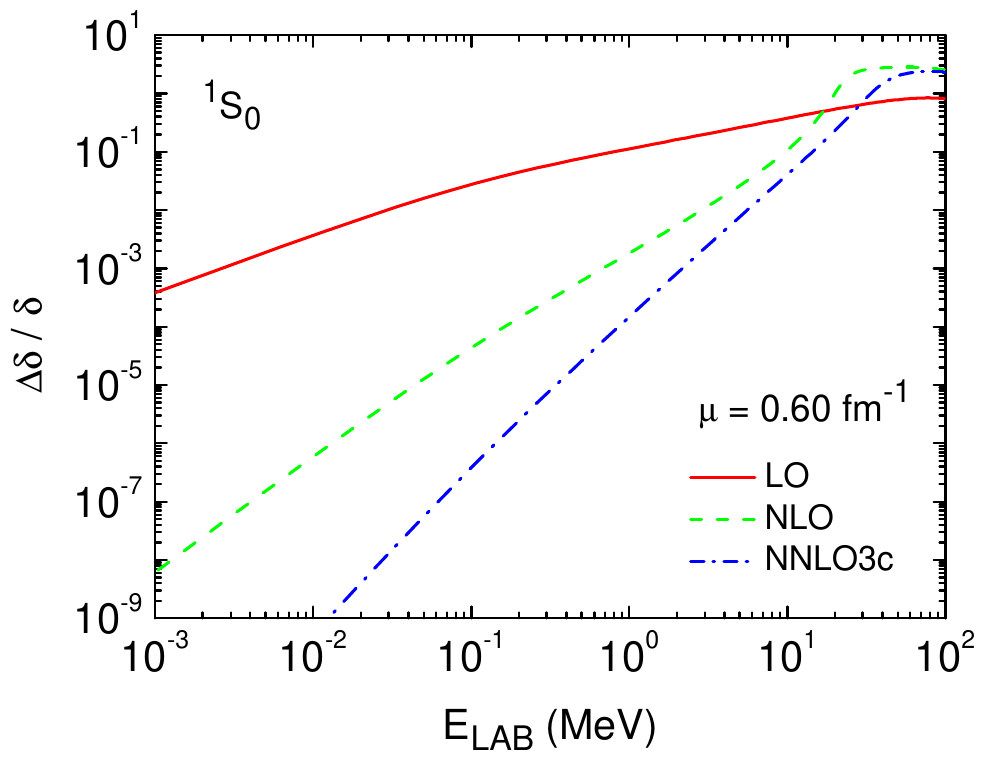}
\end{center}
\caption{(Color online) Phase-shifts in the $^1 S_{0}$ channel (left) and the corresponding relative errors (right) for the LO, NLO and NNLO potentials renormalized through the SKM procedure.}
\label{fig1}
\end{figure}

In the left panel of Fig. \ref{fig1} we show the $NN$ phase-shifts in the $^1 S_{0}$ channel obtained for the LO, NLO and NNLO potentials renormalized through the SKM approach. The strengths of the contact interactions were fixed at $\mu=0.6~{\rm fm}^{-1}$ by matching the parameters of the effective range expansion (ERE) to order $k^4$ (scattering length $a=-23.7~{\rm fm}$, effective range $r_e=2.7~{\rm fm}$ and curvature $v_2=0.48~{\rm fm}^3$), which provides a good fit to the results from the Nijmegen PWA up to on-shell momenta $k$ of order $m_{\pi} \sim 140$ MeV. In the right panel we show the log-log plots for the relative errors (with respect to the results from the ERE), where one can observe the expected order-by-order power-law improvement.

In order to analyze the renormalization group invariance of the SKM approach, we evolve the driving term $V^{(3)}$ through the NRCS equation from a reference scale ${\bar \mu}=0.76~\rm{fm}^{-1}$ to a final scale $\mu=0.63~\rm{fm}^{-1}$. 

\begin{figure}[h]
\begin{center}
\includegraphics[scale=0.33]{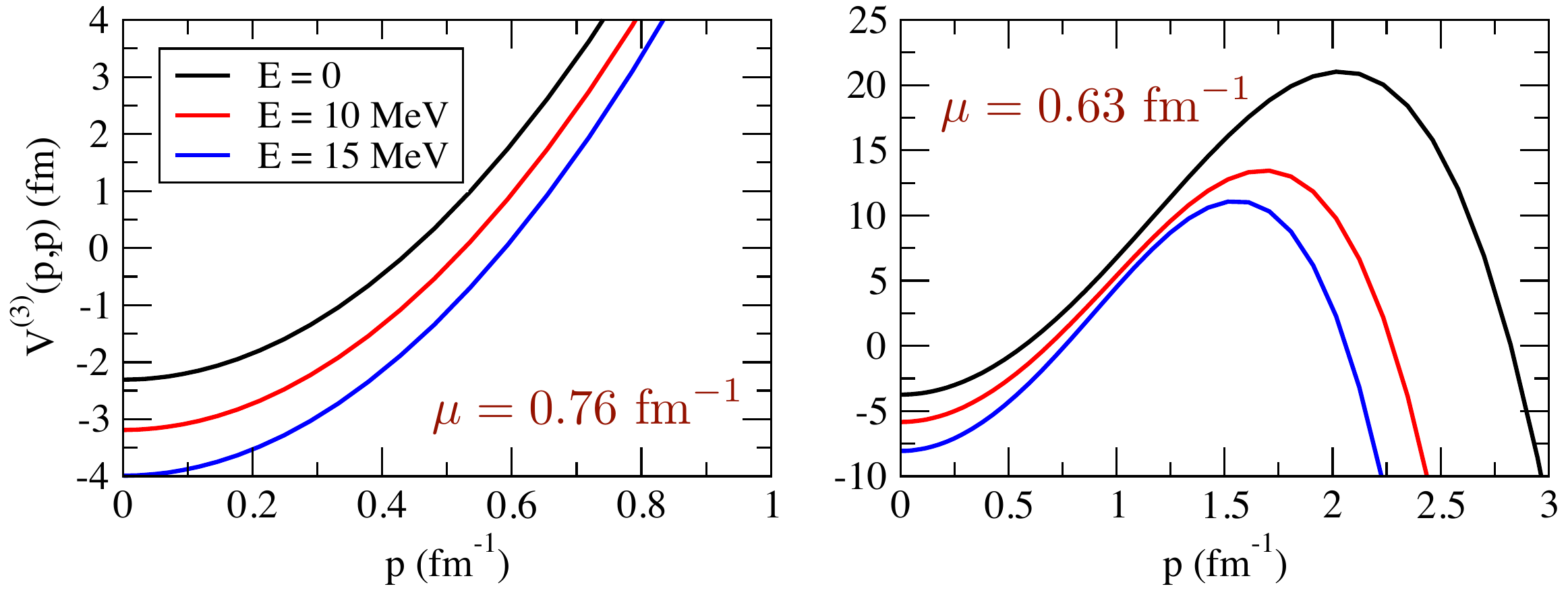}
\hspace{0.1cm}
\includegraphics[scale=0.16]{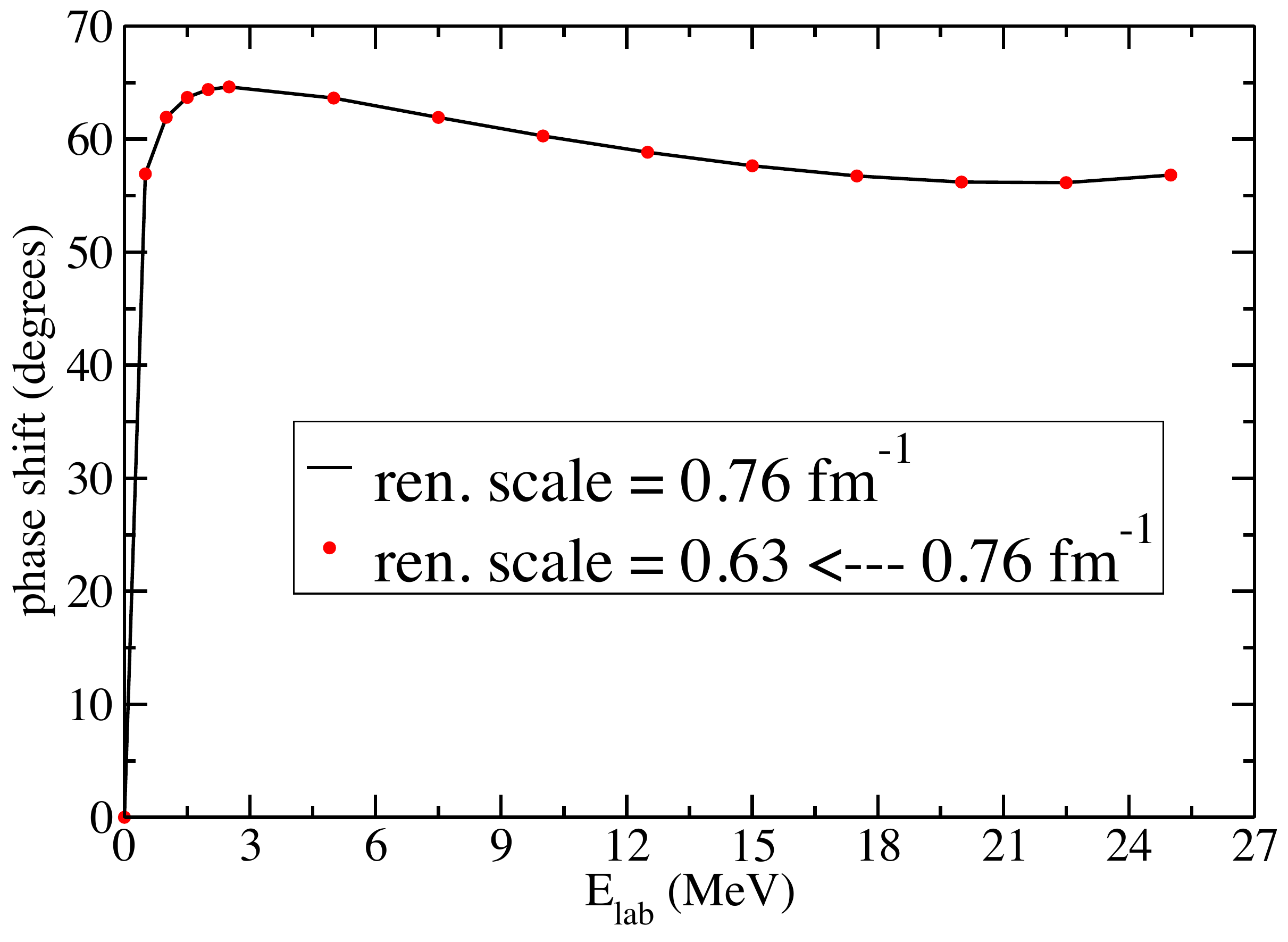}
\end{center}
\caption{(Color on-line) Evolution of the driving term $V^{(3)}$ through the NRCS equation from a reference scale ${\bar \mu}=0.76~\rm{fm}^{-1}$ (left panel) to a final scale $\mu=0.63~\rm{fm}^{-1}$ (middle panel). Phase-shifts  for the initial and the NRCS evolved driving terms (right panel).}
\label{fig2}
\end{figure}

In Fig. \ref{fig2} we show the diagonal matrix elements for the initial (left) and the NRCS evolved (middle) driving terms for $E=k^2=0,10,15$ MeV, and the phase-shifts in the $^1 S_{0}$ channel for the NLO potential calculated from the solution of the $3$-fold subtracted LS equation with the initial and the NRCS evolved driving terms (right). As one can observe, in order to ensure the invariance of the $K$-matrix under the change of the renormalization scale, the driving term significantly evolves and the results remain invariant (up to relative differences smaller than $~10^{-13}$ due to numerical errors).

\section{Conclusions}
\label{conc}

We have shown that the SKM approach with multiple subtractions provides a fully renormalization group invariant framework for treating effective $NN$ interactions in pionless EFT. In forthcoming works we will investigate the renormalization group invariance of the SKM approach in ChEFT.
\begin{acknowledgements}
This work was supported by FAPESP, FAEPEX, CNPq and Instituto Presbiteriano Mackenzie through Fundo Mackenzie de Pesquisa.
\end{acknowledgements}


\begin{thebibliography}{99}
\bibitem{epelbaum} E. Epelbaum, H.-W. Hammer and U.-G.~Mei{\ss}ner, Rev. Mod. Phys. {\bf 81}, 1773 (2009)
\bibitem{wilson1} K. G. Wilson and J. B. Kogut, Phys. Rep. {\bf 12 C}, 75 (1974)
\bibitem{wilson2} K. G. Wilson, Rev. Mod. Phys. {\b 55}, 583 (1983)
\bibitem{lepage} G. P. Lepage, nucl-th/9706029, (1997)
\bibitem{npa99} T. Frederico, V. S. Tim\'oteo and L. Tomio, Nucl. Phys. A 653, 209 (1999)
\bibitem{plb00} T. Frederico, A. Delfino and L. Tomio, Phys. Lett. B 481, 143 (2000)
\bibitem{hepph01} T. Frederico, A. Delfino, L. Tomio and V. S. Tim\'oteo, hep-ph/0101065, (2001)
\bibitem{plb05} V. S. Tim«oteo, T. Frederico, L. Tomio, and A. Delfino, Phys. Lett. B 621, 109 (2005)
\bibitem{lc09} V. S. Tim\'oteo \textit{et al.}, Nucl. Phys. B 199, 197 (2010)
\bibitem{fb19} V. S. Tim\'oteo, S. Szpigel and F. O. Dur\~aes, EPJ Web of Conf. 3, 05020 (2010)
\bibitem{iwara09b} S. Szpigel, V. S. Tim\'oteo and F. O. Dur\~aes, Int. J. Mod. Phys. D 19, No. 8-10, 1679 (2010)
\bibitem{aop2010} S. Szpigel, V. S. Tim\'oteo and F. O. Dur\~aes, doi:10.1016/j.aop.2010.11.007, (2010)
\end{thebibliography}
\end{document}